\documentclass[a4paper, twocolumn] {article}

\usepackage{xcolor}
\usepackage{graphicx}
\usepackage[superscript]{cite}
\usepackage[margin=0.585in]{geometry}
\usepackage{textcomp}
\usepackage{amsmath}
\usepackage{comment}
\usepackage{url}
\usepackage{float}
\usepackage[hidelinks]{hyperref}
\restylefloat{table}

\usepackage{pbox}
\usepackage{array}

\newcommand{\note}[1]{#1}

\title{Dielectric slot-coupled half-Maxwell fisheye lens as octave-bandwidth beam expander 
for terahertz-range applications}

\author{Daniel Headland,\footnote{Corresponding author}~\footnote{
	Graduate School of Engineering Science, Osaka University, Osaka 560-8531, Japan
	}~ 
Andreas K.~Klein,\footnote{
Department of Optoelectronics, University Duisburg-Essen, 47057 Duisburg, Germany
	}~
Masayuki Fujita,\footnotemark[1]~\footnotemark[2]~~and 
Tadao Nagatsuma,\footnotemark[2]}

\begin{document}

\maketitle

\begin{abstract}
	\textbf{
	We present a paradigm for integrated photonic devices based upon broadband 
	slab-confined collimated beams that are launched with half-Maxwell fisheye lenses. 
	Although it is challenging to match to the low-index focus of the lens whilst
	maintaining 	adequate field confinement for a well-defined point source, 
	integrated dielectric 	slot waveguides	prove highly suitable, 
	yielding collimators of 90\% efficiency and over one octave bandwidth. 
	Terahertz technology will benefit from such broadband slab-confined 
	beams to replace free-space optics, toward compact, mass-producible  systems that 
	do not require manual optical alignment.
	We present two prototype systems to demonstrate the versatility of this concept, 
	namely a diagonally-set distributed Bragg reflector as frequency-division 
	diplexer for terahertz communications, and 
	an attenuated total reflection-based liquid sensor. 
	Both are enabled by oblique in-slab reflections that are collected at a location
	other than the originating lens, which is not attainable using ordinary single-mode 
	channel waveguides. 
	}
\end{abstract}

\section{Introduction}

A dielectric slab waveguide supports infinite in-plane propagating modes, 
 and although this presents complexities in comparison to single-mode channel waveguides, 
 there are also several advantages. 
The fundamental TE$_0$ slab-mode exhibits no cutoff or leakage 
at lower frequencies, and hence it is highly broadband and 
low-dispersion.\cite{saleh2019fundamentals}
With in-slab beamforming techniques and integrated optics, this mode can 
propagate as a well-defined collimated beam.\cite{headland2020half}
The fact that these beams are free to propagate in in-plane directions allows for 
diagonally-set features to generate oblique reflections, and thereby direct the scattered
radiation away from its originating direction. 
This is in contrast to channel waveguides, for which in-guide reflections always 
propagate back towards the source. 
Thus, dielectric slab waveguides present an effective means to separate 
forward- and backward-traveling waves in broadband. 
Furthermore, overlapping beams may be separated and collected independently
if they propagate in distinct directions, as in the case of arrayed 
waveguide gratings.\cite{takahashi1992polarization}

The planar half-Maxwell fisheye lens is a semicircular gradient-index (GRIN) optic 
that operates as a beam expander by interfacing a point source at the apex 
of its circumferential arc to a collimated beam that is projected 
normally from the bisecting line.\cite{lawrence1992beam, fuchs2008comparative}
As its focal length is half its diameter, a half-Maxwell fisheye lens is innately compact. 
This manner of GRIN lens is amenable to integrated photonics, as it may be implemented
in a dielectric slab using effective medium techniques, i.e.~as
an array of subwavelength-pitch 
through-holes.\cite{hunt2012planar,headland2018terahertz,headland2020half}
If the slab is left unpatterned on the opposing side of the bisecting line, 
then the resulting structure is capable of both launching and receiving a 
slab-confined collimated beam. 
An integrated channel waveguide may interface with the lens' focus, but 
the particular choice of waveguide type is critical, as it must be matched to the 
relatively low index at the lens' circumference 
whilst avoiding delocalization of modal fields into 
the surrounding space.\cite{almeida2003nanotaper}
Previously, we have employed photonic crystal waveguides for this 
purpose,\cite{headland2020half}
but it was found that the photonic crystal adjacent to a receiving lens'
focus strongly reflected un-collected stray fields back into the slab, leading to 
undesired interference.
Here, we show that a dielectric slot waveguide, which confines
radiation within a narrow air gap,\cite{almeida2004guiding} is a highly effective solution. 
The result is  an efficient slab-mode beam collimator with over one octave bandwidth, 
outperforming integrated beam expanders 
that are based upon optimized-profile waveguide 
 tapers.\cite{abbaslou2017ultra,michaels2018leveraging}

Our experimental investigation targets the terahertz range, which will benefit from
this integrated optics paradigm.
Terahertz waves hold potential for attractive applications such as 
 imaging, \cite{kawase2003non, guerboukha2018toward} spectroscopy, 
\cite{davies2008terahertz,baxter2011terahertz}
and communications.\cite{nagatsuma2016advances,yu2016160}
However, the majority of reported demonstrations incorporate 
bulky free-space optics that require manual assembly and precise alignment, 
reducing the viable scope of real-world applications. 
As an alternative, we wish to monolithically integrate all optics of a given system within 
a single high-resistivity float-zone intrinsic silicon wafer, and fabricate them together 
in their appropriate alignment. 
In this way, we devise  two practical use-cases; a distributed Bragg
reflector (DBR)-based broadband frequency-division diplexer for terahertz communications
applications, and an attenuated total internal reflection (ATR)-based terahertz liquid 
sensor.

\section{Results}

\subsection{Lens design}
\label{design}

A Maxwell fisheye lens is described by a radially symmetrical distribution of refractive
index given by,\cite{maxwell1890scientific}
\begin{equation}
	\label{eq:maxwell}
	n(r) = \frac{n(0)}{1+\frac{r}{r_\mathrm{max}}^2},
\end{equation}
where $r_\mathrm{max}$ is the radius of the optic and $n(0)$ is the peak refractive 
index that is found at the center of the lens.
The index distribution must vary continuously between 
$n(0)$ and half this value, as $n(r_\mathrm{max})=n(0)/2$, and hence it is a requirement of 
the GRIN medium that the achievable range of refractive index must span a two-to-one ratio. 
In this work, an effective medium composed of a triangular lattice of subwavelength 
through-holes in a silicon slab provides engineerable effective index 
that is mediated by hole diameter.\cite{headland2018terahertz, gao2019effective, headland2020half, headland2020unclad,gao2020characteristics}
Our investigation targets frequencies in the vicinity of 300~GHz, for which hole pitch 
of $\sim$88~\textmu{}m is appropriate. 
In consideration of realistic fabrication constraints, the minimum viable 
hole diameter is $\sim$20~\textmu{}m, and this informs our selection of $n(0)=2.9$.
Having defined this parameter, Eq~\ref{eq:maxwell} provides a complete description of 
index distribution for the TE$_0$ slab mode that is employed in the present work. 
This translates\cite{saleh2019fundamentals} to a bulk refractive index that varies 
from $\sim$3.3 in the lens center to $\sim$1.8 at the edge, corresponding to hole diameters 
$\sim$20~\textmu{}m and $\sim$75~\textmu{}m, respectively. 
The design procedure to determine the required layout of hole sizes in the lens body
is described in detail in the Methods section. 

\begin{figure}[!b]
	\includegraphics[]{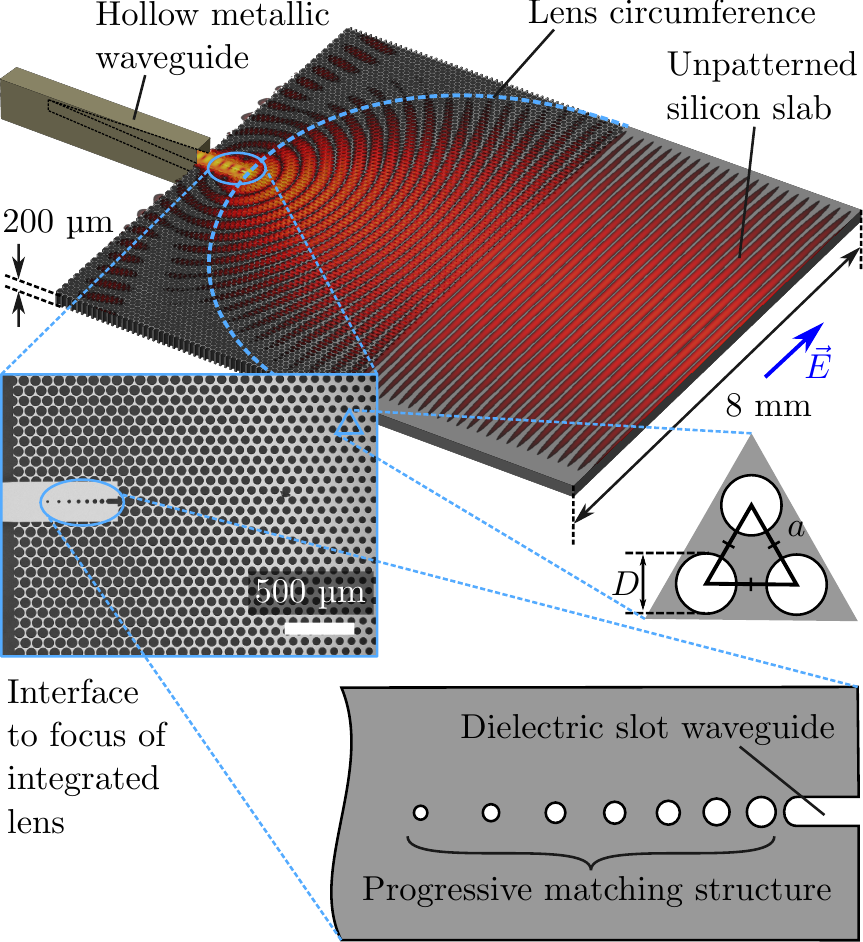}
	\caption{\label{fig:single}
	Integrated half-Maxwell fisheye lens-based slab-mode beam launcher, 
	showing a schematic diagram of the lens itself, coupling terahertz waves
	from an external hollow metallic waveguide to an un-patterned dielectric slab.
	An inset micrograph shows the detail of the feed structure that interfaces 
	to the lens' circumference, including a slot waveguide that is matched to 
	a wide channel waveguide via a progressive hole-array structure. 
	A single period of a triangular lattice of cylindrical holes is shown.
	}
\end{figure}

The optic that is the main subject of this work is illustrated in Fig.~\ref{fig:single}. 
A Maxwell fisheye lens is bisected in order to yield a half-Maxwell fisheye lens, 
and an unpatterned silicon slab is attached to the bisecting line. 
An effective medium-clad  dielectric channel 
waveguide\cite{gao2019effective,gao2020characteristics}
interfaces with the focus 
that is situated at the apex of the semicircular arc, and terahertz waves are
provided via an external hollow waveguide. 
A 2.8~mm linear-tapered spike at the termination of the channel waveguide is inserted
into the hollow metallic waveguide in order to perform broadband index matching, 
as is commonplace for experimental characterization of all-dielectric terahertz waveguide 
devices.\cite{tsuruda2015extremely,hanham2015dielectric,hanham2017led,headland2018terahertz,headland2019bragg,  gao2019effective, headland2020half, headland2020unclad,gao2020characteristics}

\begin{figure}[!tb]
	\includegraphics[]{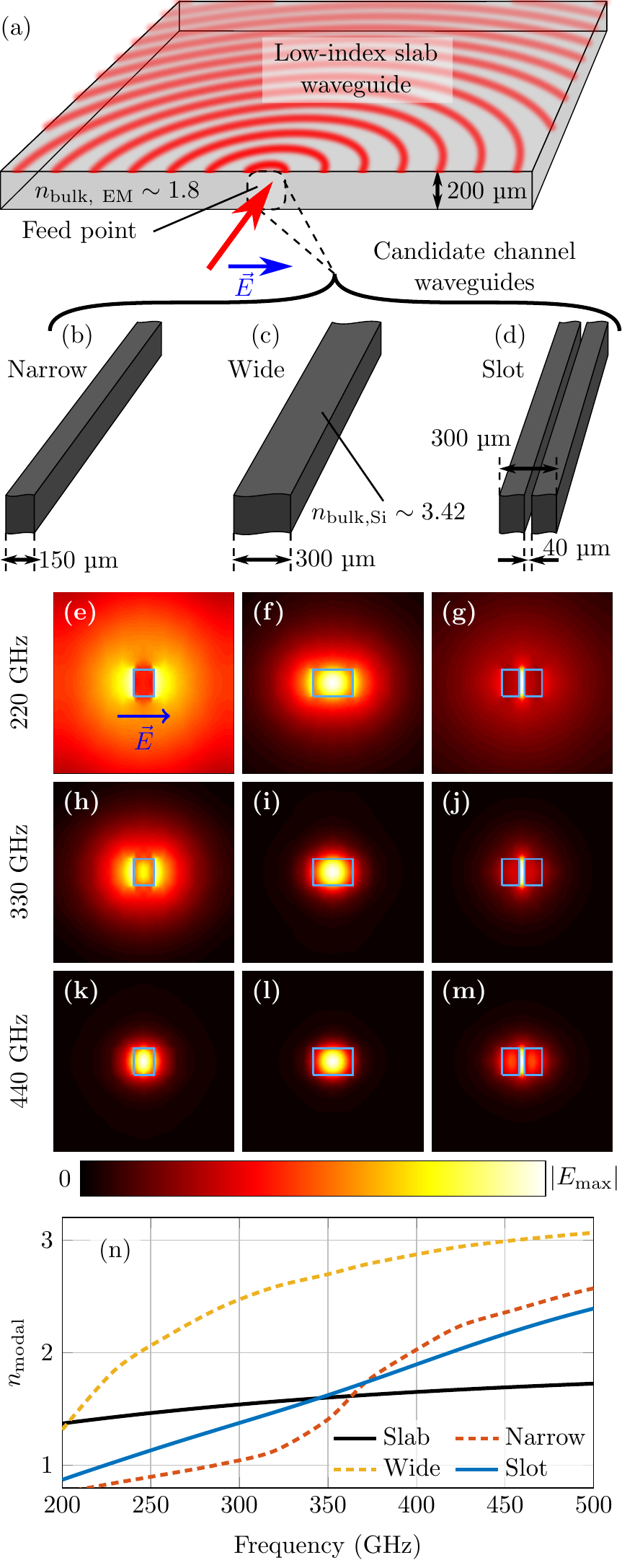}
	\caption{\label{fig:matching}
	Interface to lens, showing abstract, schematic representation of 
	(a) the low-index slab waveguide at lens' feed point, and 
	(b)--(c) candidate channel waveguides, as well as
	(e)--(m) electric field magnitude distributions of fundamental propagating modes
	of each such channel waveguide, given in linear scale and 
	normalized to their respective maxima, and 
	(n) modal indices. 
	In all cases, we consider the fundamental TE mode exclusively. 
	}
\end{figure}

The feeding structure that is employed to interface to the integrated optic 
is of critical importance, as it must provide index matching whilst 
simultaneously confining terahertz waves into a narrow point source.
For the lens, the focus may be represented as a moderately low-index dielectric slab 
region, as shown in Fig.~\ref{fig:matching}(a).
For the channel waveguide, it can be seen in the inset to Fig.~\ref{fig:single} 
that a narrow slot is etched into its termination, thereby forming  
a dielectric slot waveguide.\cite{almeida2004guiding}
This choice can be understood by  comparison with more
conventional dielectric waveguides, as shown in Figs.~\ref{fig:matching}(b),(c), 
alongside a dielectric slot waveguide in Fig.~\ref{fig:matching}(d). 
Cladding is omitted for simplicity and clarity in this discussion.
Simulated modal field distributions of candidate channel waveguides are given in 
Figs.~\ref{fig:matching}(e)--(m), and Fig.~\ref{fig:matching}(n) shows their dispersion 
relations along with that of the slab waveguide\note{, where it is noted that 
index less than unity corresponds to leaky modes that exhibit progressive loss via 
radiation to free-space}. 
It can be seen from Fig.~\ref{fig:matching}(n) that the modal index of the narrow dielectric 
waveguide given in Fig.~\ref{fig:matching}(b) is sufficiently low for reasonable index 
matching to the slab, however Figs.~\ref{fig:matching}(e) and (h) indicate that 
this is due to de-localization of modal fields into the surrounding space, which  
increases at lower frequencies. 
An undesired consequence of this is that the fields that are supplied by this waveguide
are a poor match to the point source to which the half-Maxwell fisheye lens is suited, 
and hence the resulting slab-mode beam will be of low quality. 
On the other hand, Figs.~\ref{fig:matching}(f) and (i) show that the wider dielectric
waveguide can achieve stronger field confinement, but this comes at the cost of 
greatly increased modal index, as observed in Fig.~\ref{fig:matching}(n), and this 
negatively impacts matching. 
Put simply, the dielectric waveguide presents an unwelcome trade-off that is mediated
by waveguide width; we may either achieve index matching or field confinement, but 
we cannot obtain both simultaneously. 

The dielectric slot waveguide that is illustrated in Fig.~\ref{fig:matching}(d) confines
terahertz waves strongly within an airgap, as shown in Figs.~\ref{fig:matching}(g)--(j). 
As a consequence, modal fields primarily occupy a low-index material, and hence the overall
effective modal index is reasonably low, as shown in Fig.~\ref{fig:matching}(n). 
Furthermore, the dielectric slot waveguide is of lower dispersion than the narrow dielectric
waveguide, as its modal field distribution varies less with respect to change in frequency. 
For these reasons, the dielectric slot waveguide is well suited to feed the 
integrated half-Maxwell fisheye lens. 

Having selected the appropriate manner of feeding structure for the integrated 
half-Maxwell fisheye lens-based slab-mode beam launcher, we may proceed with the 
details of the feed design. 
Although we employ a hollow metallic waveguide to provide an interface to the external 
world, it is unsuited to launch a dielectric slot mode directly, and hence 
a 306~\textmu{}m-wide dielectric waveguide is employed as an interface. 
A progressive hole-matching structure serves as a compact transition between the 
dieletric waveguide and dielectric slot modes. 
There are seven holes, of diameter varying from 20~\textmu{}m to 35~\textmu{}m, 
and hole separation  is from 63~\textmu{}m to 13~\textmu{}m. 
The slot itself is 37~\textmu{}m-wide and 70~\textmu{}m-long. 
This final design is shown in Fig.~\ref{fig:single}.

\begin{figure}[!tb]
	\includegraphics[]{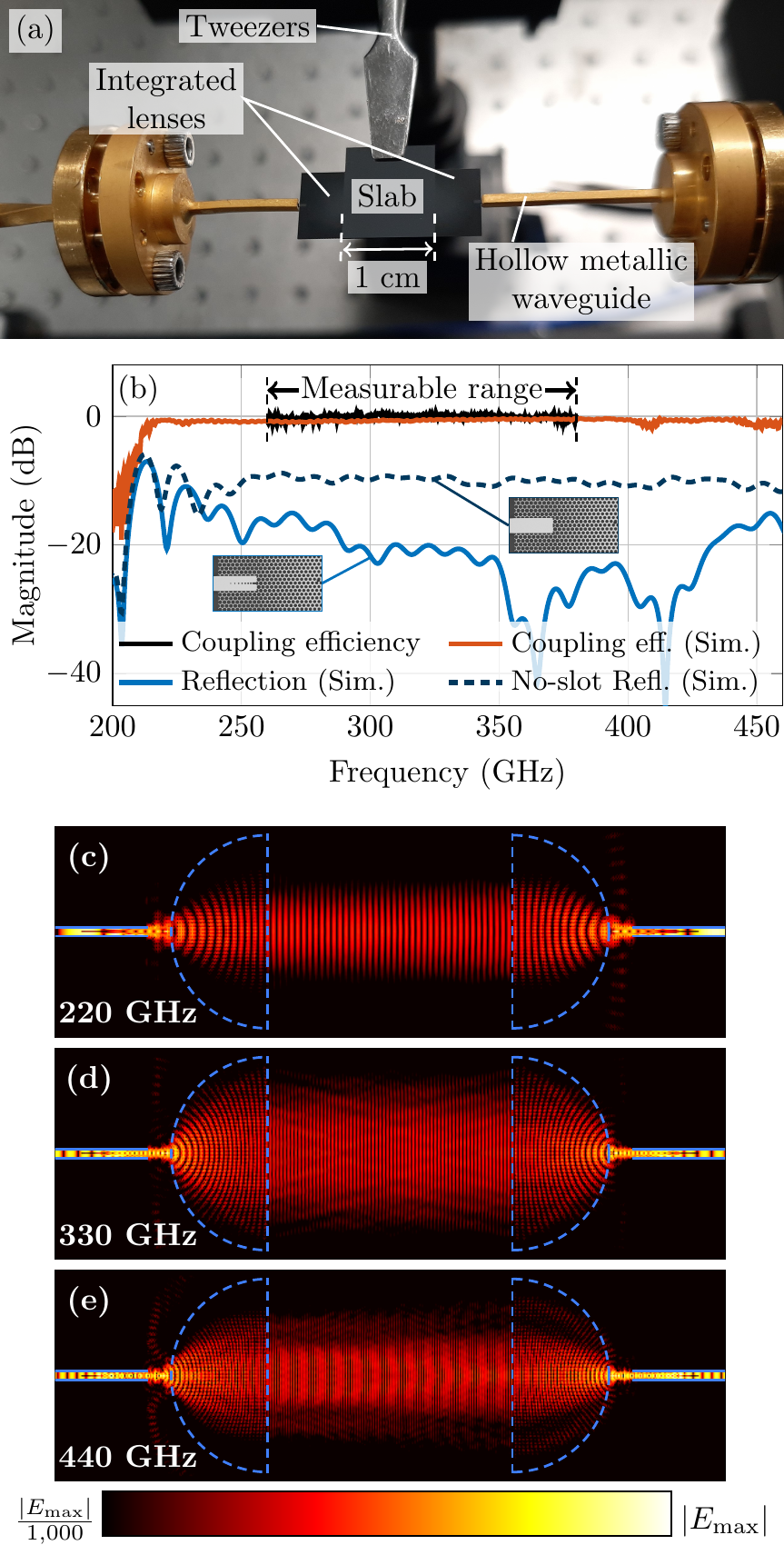}
	\caption{\label{fig:twolens}
	Evaluation of the half-Maxwell fisheye lens' performance by means of 
	a two-lens sample, which is shown undergoing experimental characterization in (a). 
	(b) Coupling efficiency of a single lens, as well as simulated matching performance
	as represented by reflection magnitude, both with and without the slot waveguide
	and associated matching structure, and 
	(c)--(e) simulated field magnitude distributions of two-lens structure at 
	several frequencies of interest, where each is shown in logarithmic scale, and 
	normalized to its respective maximum. 
	}
\end{figure}

In order to experimentally verify the existence of the slab-mode beam, we devise a 
sample that consists of two lenses that face each other across a 1~cm-long slab 
region as a simple transmission configuration.
A source is coupled to one of the lenses and a detector is coupled to the other, 
and hence successful transfer of terahertz power through the sample constitutes
evidence for the existence of the slab-mode beam. 
This experiment is shown in the photograph in Fig.~\ref{fig:twolens}(a). 
The coupling efficiency of a single lens is extracted from the measured 
transmission, and is given in Fig.~\ref{fig:twolens}(b). 
It can be seen that the 3-dB bandwidth of the optic is not encountered across
the measurable range. 
Fortunately, the frequency range of full-wave simulations is less constrained, and
hence simulations facilitate a broader-bandwidth investigation.
Such techniques also provide insight into lens matching performance 
and field distributions, which are not measurable with our available equipment.
Simulated results given in Fig.~\ref{fig:twolens}(b) show that the lower-frequency 
cutoff is encountered at $\sim$216~GHz, and above this frequency, reflection magnitude 
is below $-$10~dB. 
Overall bandwidth exceeds one octave, and peak transmission efficiency is $\sim$90\%, 
but it is noted that this performance corresponds to ideal alignment with the 
hollow waveguide.
Simulated field distributions are given in Figs.~\ref{fig:twolens}(c)--(e), where it 
can be seen that a high-quality collimated beam is produced over a broad bandwidth. 
Insight is sought into the impact of the dielectric slot waveguide and associated matching 
structure, and hence these components are removed, leaving a wide dielectric waveguide, 
and reflection response is re-calculated via full-wave simulation. 
This scenario is illustrated inset to Fig.~\ref{fig:twolens}(b), and it can be seen that 
reflection magnitude has indeed increased to $\sim$$-$10~dB across a broad bandwidth.

\subsection{Prototype systems and applications}
\label{applications}

Having completed the design of the integrated terahertz lens in Section~\ref{design}, 
the aim of the present section is to develop prototype integrated optical systems in order 
to demonstrate the practical utility of the emerging terahertz slab-mode beam paradigm. 
A range of passive functional optical components may be implemented directly in the 
slab using an arrangement of through-holes, and 
integrated beam splitters are one such possibility. 
To this end, a stripe of effective medium can produce an index step, and thereby generate
reflections of magnitude that is dependent upon air filling-factor and the width of the 
stripe. 
Crucially, the use of freely propagating beams allows this stripe to be set obliquely, 
such that the reflected radiation may be directed to a location other than that of its 
originating lens. 
Thus, the slab-mode beam paradigm can separate backward- and forwards-traveling waves in 
a manner that linear channel waveguides cannot.

\begin{figure}[!tb]
	\includegraphics[]{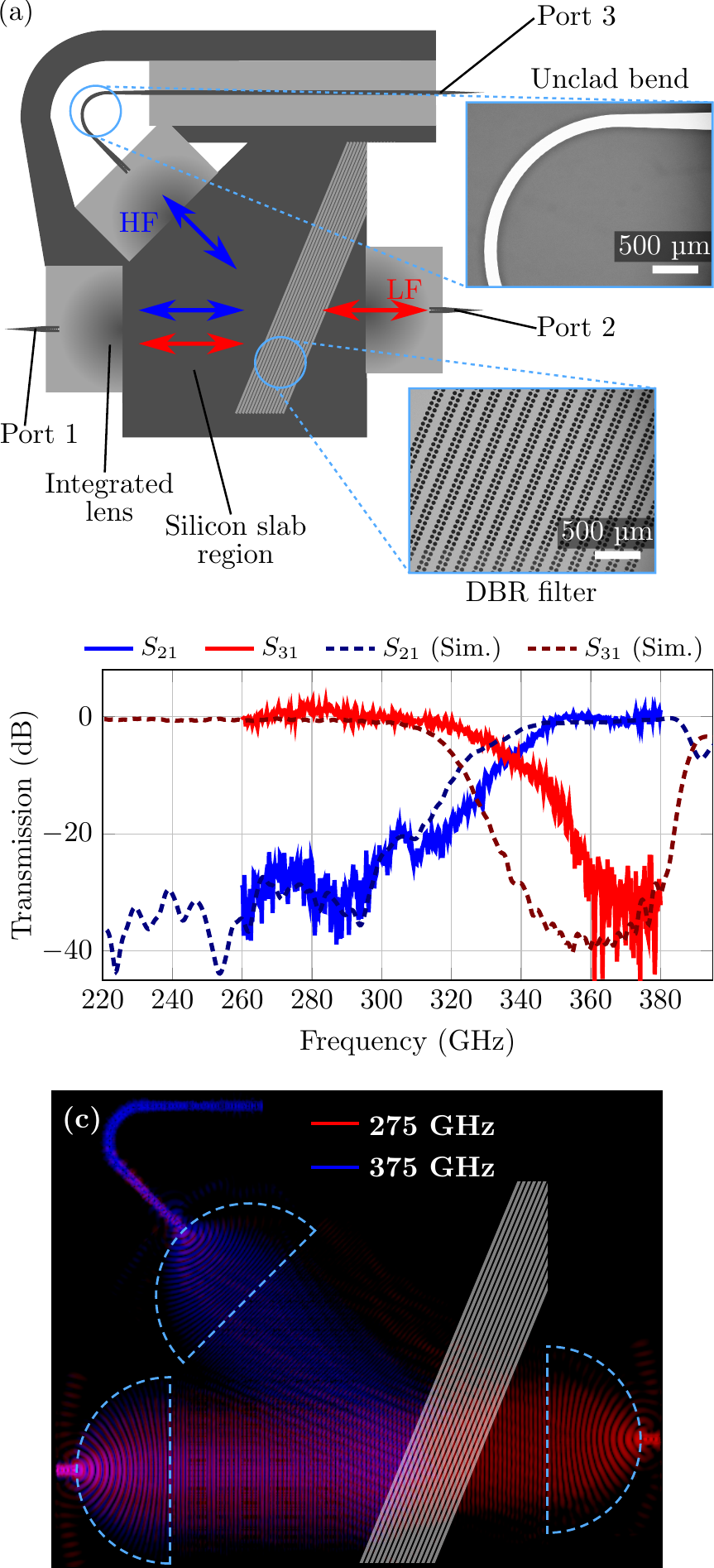}
	\caption{\label{fig:DBR}
	Integrated DBR-based frequency diplexer, showing 
	(a) schematic, with inset micrographs, 
	(b) transmission magnitude between each pair of ports, and
	(c) simulated field magnitude distributions at two frequencies of interest, in
	false-color representation. 
	Each field plot is normalized to its own maximum, and is given in logarithmic 
	scale with 30~dB dynamic range. 
	}
\end{figure}

It is well understood that periodic alternation of a medium's dielectric properties
can produce frequency-dependent transmission and 
reflection.\cite{vincent1994optical,diener2001dichroic}
This concept may be implemented in a slab waveguide by means of an array of 
 stripes of effective medium, thereby realizing a DBR or dichroic beam splitter, 
 which serves as an optical filter in the present work.
As illustrated in Fig.~\ref{fig:DBR}(a), such a structure may be integrated directly 
with a three-lens setup in order to realize an all-dielectric broadband frequency-division 
diplexer. 
The DBR optical filter is set at a 22.5$^\circ$ angle, in order to achieve 45$^\circ$ 
separation between the originating lens and that which collects the filter's stopband. 
The filter is optimized for cross-over at $\sim$320~GHz, \note{i.e.~the center of 
our measurable range,} and is composed of 47~\textmu{}m-diameter cylindrical through-holes 
in a 64~\textmu{}m-period square lattice. 
Each stripe constitutes two rows of through-holes, and the periodic separation between
adjacent stripes is 203~\textmu{}m. 
It is noted that, although this microstructure is effective medium, a single stripe 
contains too few rows of holes to be accurately modeled using effective-medium theory, 
for which approximate local uniformity is required. 
Thus, it is not possible to provide a meaningful value for effective refractive index. 
There are fifteen stripes in total, and a micrograph of the integrated filter is shown inset 
to Fig.~\ref{fig:DBR}(a). 
The lens that is intended to collect the passband is raised by 1.3~mm with respect to the 
originating lens, in order to account for refraction during transit through the 
lower-effective index portions of the optical filter. 
A 135$^\circ$ unclad bend\cite{headland2020unclad} is employed to re-route the reflected 
stopband following transit through its associated lens, such that it exits the sample  
parallel to the passband port in order to facilitate ease of probing. 
In addition to this, the bend structure's width is narrowed to 145~\textmu{}m via
linear tapers in order to radiate low frequencies and improve out-of-band rejection. 
The principle of operation of this bend-as-filter structure is that narrow waveguides 
exhibit progressively weaker field confinement with respect to decrease in operation 
frequency, as shown in Fig.~\ref{fig:matching}(e),(h), and (k).
A micrograph of the bend is given as inset to Fig.~\ref{fig:DBR}(a).

The measured transmission of the integrated DBR-based terahertz frequency-division
diplexer is given in Fig.~\ref{fig:DBR}(c), along with simulated response. 
It can be seen that efficient broadband frequency-splitting is indeed achieved, 
and hence the intended functionality of this device is validated. 
However, the crossover-point has shifted upwards in frequency by $\sim$18~GHz. 
We ascribe this to the employed wafer thickness being lower than its nominal 
specification, leading to changes in the dispersion profile of the slab and 
effective medium that comprise the integrated filter. 
A thinner wafer will also produce over-etching of the small holes of the filter, which 
reduces the effective index of the stripes, and increases the frequency of the wavelength
that corresponds to a stopband period. 
False-color field plots from the full-wave simulations are given in Fig.~\ref{fig:DBR}(c), 
showing the separation of the slab-confined beam into two distinct frequencies by the 
integrated DBR-based optical filter. 

\begin{figure}[!tb]
	\includegraphics[]{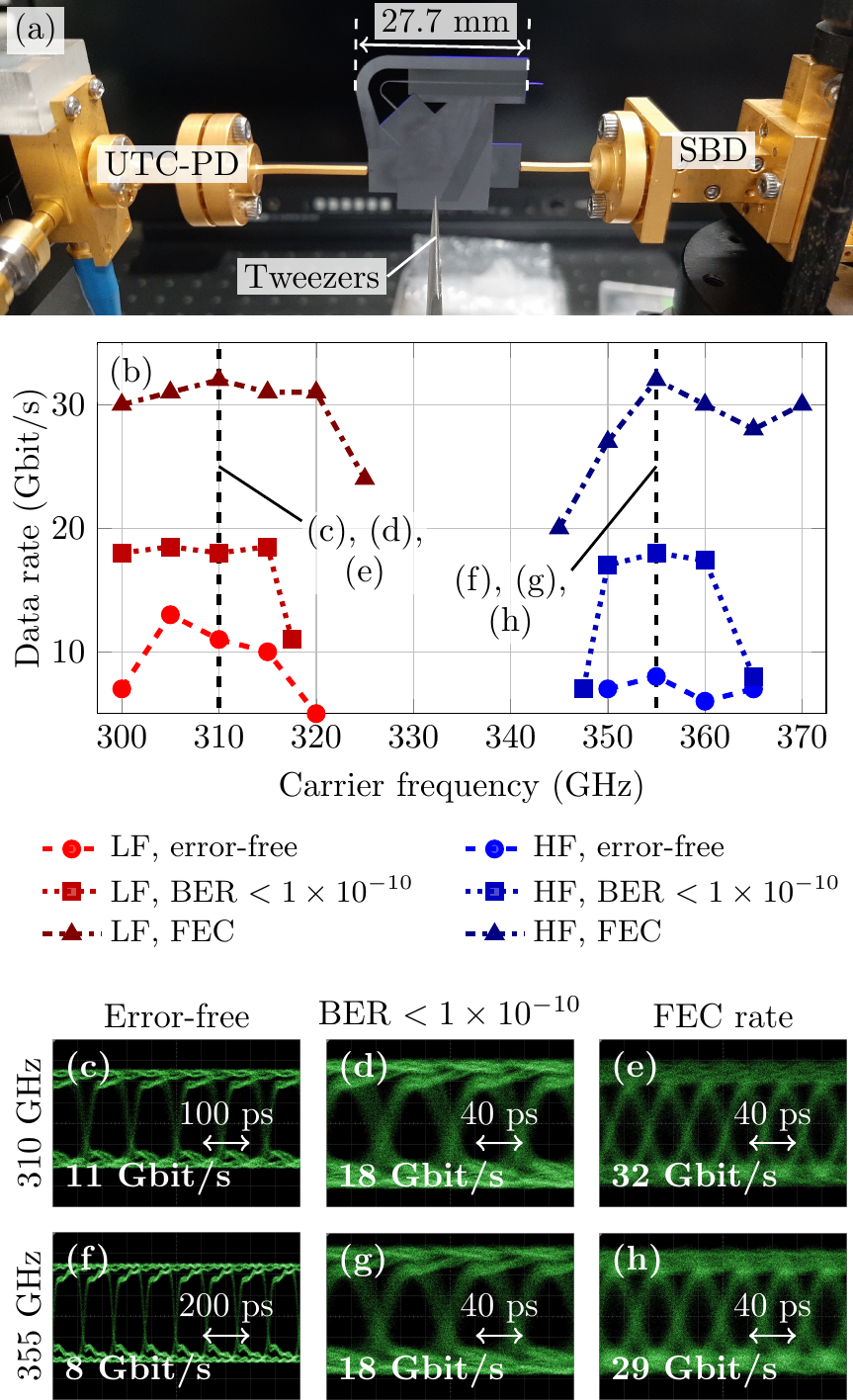}
	\caption{\label{fig:comms}
	Demonstration of terahertz communications, showing 
	(a) photograph of experiment, 
	(b) achievable data rates at three different maximum
	bit-error rates, for a variety of carrier frequencies, 
	and 
	(c)--(h) eye diagrams. 
	}
\end{figure}

The primary intended application of the frequency-division diplexer is high-volume 
terahertz-range communications,\cite{nagatsuma2016advances} 
and hence it is desirable to validate its applicability thereto. 
To this end, on-off-keying-modulated terahertz waves are generated using a uni-traveling 
carrier photodiode (UTC-PD), conveyed through the integrated diplexer, and subsequently 
detected by a Schottky-barrier diode (SBD). 
A photograph of this experiment is included in Fig.~\ref{fig:comms}(a). 
Both channels of the diplexer are employed, and a range of terahertz 
carrier frequencies within their respective transmission bandwidths is utilized. 
In each case, the highest data rate is found that abides a given maximum measured 
bit-error rate (BER) within a finite 30-second timespan.
The three chosen maximum BER values are ``error-free,'' for which no errors are counted 
(although finite errors would be counted if the timespan were 
extended indefinitely), $\mathrm{BER}<1\times10^{-10}$, and ``forward error-correction 
(FEC)''-achievable rate, which is $2\times10^{-3}$.\cite{union2004itu} 
The results of this experiment are given in Fig.~\ref{fig:comms}(b). 
It can be seen that data rates of between $\sim$8~Gbit/s and $\sim$30~Gbit/s
are achievable, where higher permitted BER is associated with increased achievable 
data rates, as expected. 
Additional to that, performance is observed to decrease as carrier frequency
approaches the cross-over frequency of the device. 
Eye diagrams at 310~GHz and 355~GHz are given in Fig.~\ref{fig:comms}(c)-(h), and
a clearly-open ``eye'' is visible in each case. 

\begin{figure}[!tb]
	\includegraphics[]{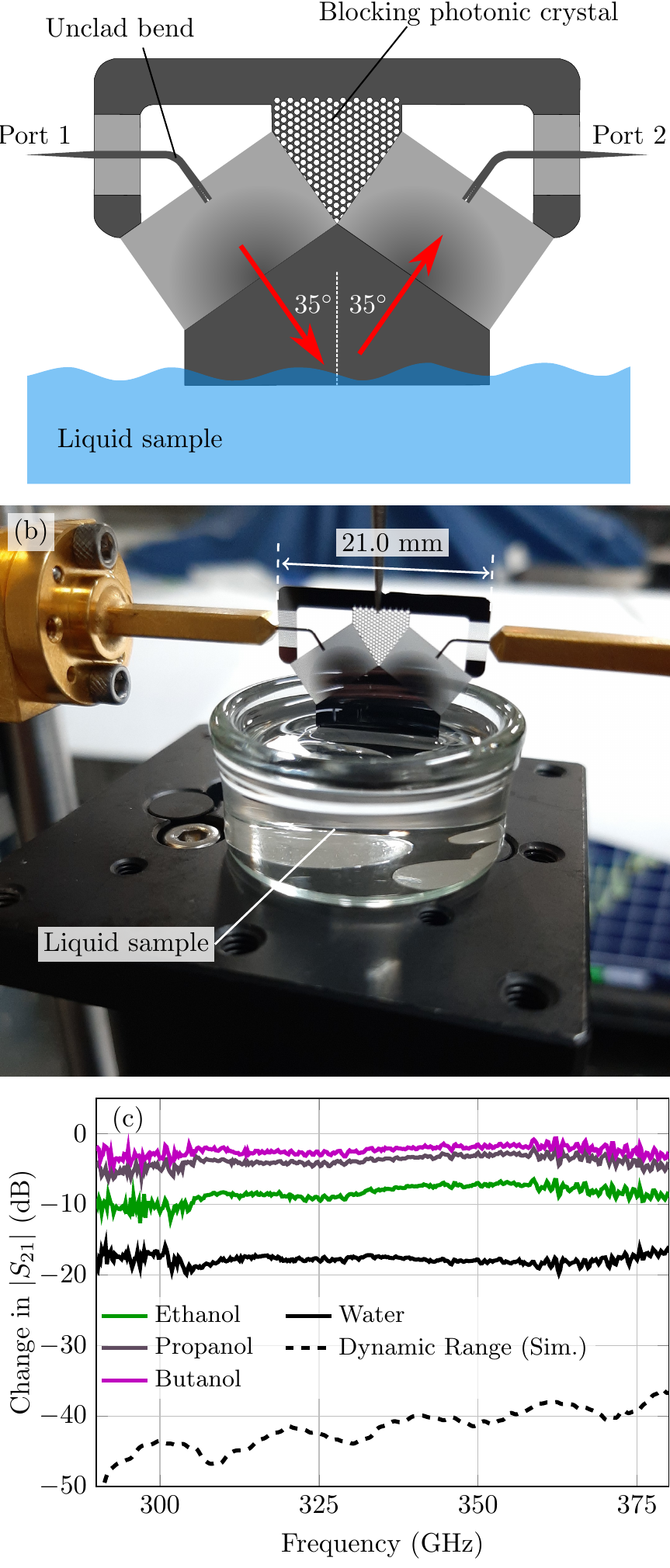}
	\caption{\label{fig:liquid}
	Integrated ATR-based dip sensor, showing
	(a) schematic representation, 
	(b) photograph of liquid sensing experiment, and
	(c) the change, or deviation in transmission magnitude due to the presence of a 
	variety of polar liquids at the sensing area, as compared to the case in which 
	only air is present. 
	}
\end{figure}

Aside from broadband communications, terahertz waves have also been nominated 
for applications in spectroscopy\cite{davies2008terahertz,baxter2011terahertz} and 
liquid sensing.\cite{hanham2015dielectric,swithenbank2017chip}
To support the integration of systems of this sort, it is possible to define sensing 
areas  in the silicon slab medium where the slab-mode interfaces with the a given analyte
in order to probe its properties.
A simple example of a sensing area is a straight edge. 
If a terahertz slab-mode beam is impinged upon this interface at an oblique angle, 
then total internal reflection (TIR) will occur due to the large index contrast 
between air and silicon. 
On the other hand, if a liquid substance of greater-than-unity refractive index is 
introduced to the sensing area then TIR may be inhibited, resulting in attenuation 
of the reflected wave. 
Thus, it is possible to sense the presence and properties of that liquid by measuring the 
attenuation of the reflected beam, thereby realizing ATR-based
sensing.\cite{nagai2006terahertz}
An integrated slab-mode device to realize terahertz ATR is illustrated in 
Fig.~\ref{fig:liquid}(a). 
Unclad bends\cite{headland2020unclad} are employed in order to feed the integrated 
lenses at the desired angle, whilst maintaining  ports that are co-linear and anti-parallel 
for ease of probing. 
As these two ports face each other directly, broadband photonic crystal medium is included 
in the space between them as a precaution in order to inhibit undesired direct transmission,
whilst providing physical support. 
An angle of incidence of 35$^\circ$ is chosen in order to maximize the dynamic range (DR)
between the refractive index of water and unity. 

The integrated ATR sample is fabricated and deployed as a terahertz dip-sensor, 
as shown in the photograph in Fig.~\ref{fig:liquid}(b). 
We employ several common polar solvents, namely distilled water and alcohols of 
different chain length, and determine the change in transmission magnitude between 
the two ports due to the presence of the liquid. 
The results of this experiment are given in Fig.~\ref{fig:liquid}(c), and 
it can be seen that all of the polar solvents employed are clearly distinguishable by their 
respective reflection. 
The measured DR of the liquid sensor is 18--20~dB, corresponding to the 
case of distilled water, which exhibited the largest overall loss. 
\note{The theoretical maximum DR of the sensor is limited by undesired transmission of 
stray terahertz power, which does not interact with the sensing area, between Ports~1 and 2. 
This is investigated in simulation by absorbing all fields that are incident upon
the sensing area, and the results given in Fig.~\ref{fig:liquid}(c)
show that theoretical maximum DR is greater than 37~dB across the measured 
bandwidth.}

As for the alcohols, a clear tendency of increasing reflection with respect to decrease in
refractive index can be observed.
This is to be expected as alcohols of greater chain length exhibit increased index 
contrast to the silicon. 
Propanol and butanol are distinguishable despite the small difference in relative 
permittivity ($\Delta \epsilon '$ and $\Delta \epsilon '' <  0.1$), thereby demonstrating 
the high sensitivity of the liquid sensor.\cite{swithenbank2017chip}


\section{Discussion}

We have realized an efficient dielectric slab-mode beam launcher with bandwidth 
over one octave.
To support this, we have shown that a dielectric slot waveguide serves as
an effective means to interface a low-index slab waveguide to a dielectric channel waveguide, and we deploy this solution to feed an integrated half-Maxwell fisheye lens. 
It is noted  that a multitude of other GRIN optics may also benefit from this technique, 
including Luneburg and Eaton 
lenses\cite{zentgraf2011plasmonic,du20163,sayanskiy2017broadband} 
across the electromagnetic spectrum. 
Aside from integrated lenses, it noted that quasi-conformal transformation optics
may also be implemented in gradient-index effective 
medium,\cite{valentine2009optical, gabrielli2010transformation} 
and hence the dielectric slot waveguide may serve to directly feed a broad range of 
advanced transformation-optics devices.

One limitation of the dielectric slot waveguide feed is its polarization. 
We have employed the TE  polarization (i.e.~$\vec{E}$-field is surface-parallel)
exclusively in this work. 
Although the TM  polarization (i.e.~$\vec{E}$-field is surface-normal) is also supported
by a dielectric slot waveguide, its modal fields are not confined within the airgap, and
so it does not exhibit comparably narrow confinement, and its dispersion relation is 
quite different. 
Thus, the slab-mode beam launcher that is the main subject of this work  will not 
support polarization-diverse applications. 
Another potential limitation is the sensitivity of micro-scale hole lattice structures to 
fabrication errors and variation in thickness between silicon wafers, as in the 
case of the DBR-based optical filter, for which the crossover frequency was shifted 
upwards by $\sim$18~GHz. 
In order for future systems of this sort to be commercially viable, it will be necessary
to develop  optical filter designs that are more robust to tolerances, or to lay out 
the photolithographic mask in such a way that compensates for anticipated over-etching. 
Nevertheless, the frequency-division diplexer serves as a useful proof-of-concept, 
and its applicability to terahertz-range communications applications was demonstrated.

We wish to remark that effective medium techniques are not the sole viable means 
to realize GRIN terahertz optics, as the dispersion of a parallel-plate metallic waveguide
may also serve to realize engineerable refractive index that is mediated by plate separation,
and this has been demonstrated for Luneburg and Maxwell-fisheye lenses.
\cite{liu2013maxwell,amarasinghe2019luneburg,sato2020terahertz} 
That said, such devices exhibit Ohmic loss, and have no potential for integration.
It is also noted that, although dielectric slot waveguides have previously been 
studied in the terahertz range,\cite{nagel2006low,amarloo2018terahertz,aller2019quasi} 
they have never been employed to directly feed optics at any frequency range. 

The capacity to efficiently launch a terahertz slab-mode beam in broadband opens the 
door to realize diverse integrated terahertz optical systems. 
In contrast to devices that depend upon free-space beams, integrated systems
will be considerably more compact, and will not require manual assembly and alignment of 
individual optics. 
Functional components, sensing areas, and other interfaces may be implemented directly
in microstructured silicon. 
We have presented two integrated devices as concrete examples of viable usage cases---a
DBR-based frequency-division diplexer and an ATR-based liquid sensor---both of which operate 
in broadband. 
It is likely that many other terahertz systems may be implemented in this way, to support
a plethora of diverse applications of terahertz waves. 
\note{Furthermore, terahertz technology will not be the sole beneficiary of broadband 
slab-mode beam launchers, as these passive, all-dielectric structures may be implemented 
across the electromagnetic spectrum, and are amenable to nanophotonics in 
particular.\cite{gabrielli2010transformation,valentine2009optical}
For instance, quantum-mechanical photonic tecnhologies frequently rely upon oblique 
reflections of individual photons from free-space beam 
splitters.\cite{o2009photonic,kutas2020terahertz}
Slab-mode techniques represent a possibility to directly implement such 
techniques in a broadband integrated platform.}

All of the structures presented in this work are implemented using through-holes in a 
silicon slab exclusively. 
This is a deliberate choice for the sake of simplicity, as only a single-mask etch process is 
required. 
On the other hand, the scope of terahertz applications would likely be expanded 
considerably by patterning a metal top layer, incorporating 
phase-change materials,\cite{taha2017insulator} or by doping of the silicon itself. 
Hybrid integration with active terahertz devices also presents an opportunity to incorporate
active terahertz devices,\cite{yu2019efficient,yu2020waveguide} 
thereby rendering the hollow-metallic-waveguide interface obsolete, and reducing 
systems size further. 

\section{Materials and Methods}

\subsection{Model generation}

\subsubsection{Arrangement of through-holes in the lens}

The half-Maxwell fisheye lens enjoys a closed-form design procedure, which we
have previously presented.\cite{headland2020half}
Firstly, it is noted that radiation propagates within the lens as a mildly-dispersive 
TE$_m$ slab-mode, and hence Eq.~\ref{eq:maxwell} is a description of slab-mode index, 
$n_{\mathrm{slab},m}$. 
This must be translated into the refractive index of a bulk material, $n_\mathrm{eff}$, 
for which a description of the dispersion relation of the slab 
waveguide is required,\cite{saleh2019fundamentals} 
\begin{equation}\label{eq:phaseconst}
	\begin{split}
\tan^2 \left\{ \frac{t}{2} \sqrt{ (n_\mathrm{eff} k_0)^2 - (k_0 n_{\mathrm{slab},m})^2} - \frac{m \pi}{2} \right\} \\= \frac{(k_0 n_{\mathrm{slab},m})^2 - k_0^2}{(n_\mathrm{eff} k_0)^2 - (k_0 n_{\mathrm{slab},m})^2}, 
	\end{split}
\end{equation}
where $k_0$ is free-space wavenumber and $t$ is dielectric slab thickness, which is 
equal to 200~\textmu{}m in this work. 
The fundamental TE mode is employed in the present work, and hence $m=0$.
Solving this transcendental equation yields the required distribution of bulk
effective index, $n_\mathrm{eff}$, as a function of radial position within the lens body. 
Finally, this is converted to a layout of hole diameters by means of the 
Maxwell Garnett effective medium approximation,\cite{subashiev2006modal}
\begin{equation}
 n_\mathrm{eff}^{2} = \epsilon_\mathrm{Si} \frac{(1 + \epsilon_\mathrm{Si}) + (1 - \epsilon_\mathrm{Si})\zeta}{(1 + \epsilon_\mathrm{Si}) - (1 - \epsilon_\mathrm{Si})\zeta},
\end{equation}
where $\zeta$ is the volumetric air filling-factor, which is calculated using the 
equilateral triangular hole-lattice geometry shown in Fig.~\ref{fig:single}, 
\begin{equation}
	\zeta = \left(\frac{D}{a}\right)^{2} \frac{\pi}{2 \sqrt{3}}.
\end{equation}

\subsubsection{Through-holes to clad channel waveguides}

Following the example of Refs.~\citen{gao2019effective, gao2020characteristics}, 
we clad the channel waveguide with effective medium that is as low-index as possible, 
and as a consequence, we wish to employ maximum viable hole diameter. 
In order to maintain structural robustness, we employ a minimum distance of 10~\textmu{}m
between adjacent holes. 
The hole lattice that comprises the lens itself is simply extended 
to form the cladding of the channel waveguide, and hence the two regions of effective 
medium share a common hole pitch.
Thus, the diameter of the waveguide-cladding air holes is 78~\textmu{}m. 

Semicircular holes make direct contact with the sides of the channel waveguide as a 
precaution against spatial index modulation along the propagation direction, which may 
cause undesired Bragg mirror effects at higher frequencies,\cite{headland2019bragg} and 
thereby reduce bandwidth. 
This necessitates that the ratio of waveguide width to $\frac{a\sqrt{3}}{2}$ must be 
an integer, in order that both edges of the waveguide bisect a row of holes. 
This is ensured by making the hole pitch dependent upon waveguide width, and scripting 
the generation of the layout of holes in the lens body to update automatically in response
to parametric adjustment of waveguide width. 
This manner of interconnected, parametric model generation facilitates convenient 
optimization. 

\subsection{Simulation}


All simulations are performed using the commercially available CST Studio Suite 
electromagnetic full-wave simulation software package.

\subsubsection{Full-wave simulation model}

The frequency-domain mode solver of CST Studio Suite is employed to determine 
the modal field distributions and dispersion relations that are given in 
Fig.~\ref{fig:matching}(e)--(n). 
Symmetry planes are employed in order to ensure that the displayed fundamental mode is
of the desired polarization. 

The simulation results given in Figs.~\ref{fig:twolens}(b)--(e) and \ref{fig:DBR}(b),(c)
are generated with the time-domain solver of CST Studio Suite, which employs the 
finite integral technique. 
In the reflection-magnitude results given in Fig.~\ref{fig:twolens}(b), 
a single-lens model is employed, and the outgoing beam is simply discarded by attaching 
the slab-region to an absorbing boundary. 
In this way, the reflections that are generated within the slab-mode beam launcher 
structure are essentially isolated, and the impact of features within the slab, 
as well as output coupling, are neglected. 

In the case of the simulations of the DBR-based frequency-division diplexer that 
is given in Fig.~\ref{fig:DBR}(b),(c), some approximations 
are made in order to reduce the electrically-large calculation domain to a practical, 
manageable scope. 
The hole-lattice arrays that comprise the lenses themselves are replaced with a set of 
concentric semi-annular rings of bulk refractive index such 
that the TE$_0$ slab-mode index distribution corresponds to Eq.~\ref{eq:maxwell}.
Furthermore, the channel waveguide that conveys the DBR filter's stopband to Port~3 is 
shortened, so as to reduce overall pulse-propagation time through the model. 

For the liquid sensor that is presented in Fig.~\ref{fig:liquid}, the undesired 
direct transmission between Ports~1 and 2 is determined by placing an absorbing 
boundary at the surface of the sensing area. 
In this way, any field that would ordinarily interact with the analyte is removed 
from the calculation domain. 
The simulated transmission is normalized against that of a simulation in which 
the sensing area is in contact with air, and the results are presented in 
Fig.~\ref{fig:liquid}(c).

\subsubsection{Optimization of microstructures}

Optimization of microstructures such as the slot-waveguide feed and progressive matching 
structure shown in Fig.~\ref{fig:matching}, as well as the DBR-based filter shown in 
Fig.~\ref{fig:DBR}(a), is performed manually. 
This consists of a repeated cycle of full-wave simulation, subjective appraisal of the 
quality of field distributions and $S$-parameters, and adjustment of the physical parameters 
that define the geometry of the structure in question. 
No automated optimization algorithm was employed. 
The reason for this choice is that the devices that are presented in this work 
are electrically large, and require a fine mesh for accurate representation of effective
medium structures. 
As a consequence, simulation time can span several days. 
The amount of time that is required for manual adjustment by the designer is 
therefore insignificant compared to the overall simulation time, and hence there 
the usage of automation does not present an advantage in terms of time-efficiency. 
This is because all optimization algorithms require a large number of simulations in order
to produce a final design, whereas an experienced user is capable of making educated
choices in each simulation step.

\subsection{Experimental methods}

\subsubsection{Fabrication}

All silicon devices are fabricated using deep-reactive ion etching (DRIE), which 
employs a single photolithographic mask. 
Several devices are etched concurrently from a single 200~\textmu{}m-thick, 10~cm-diameter
high-resistivity float-zone intrinsic silicon wafer. 

\subsubsection{Handling} 

Silicon samples are handled using ordinary tweezers that make contact with 
unpatterned portions of the device that serve no electromagnetic purpose. 
When the sample is secured for probing, the tweezers that bear the sample are 
manually clamped into a vice. 

\subsubsection{Characterization of transmission magnitude}

Terahertz waves are generated electronically, via up-conversion of a millimeter-wave 
signal with a $\times$9 multiplier, and then output by a hollow metallic waveguide 
of rectangular internal dimensions $711\times356$~\textmu{}m (i.e.~WR-2.8). 
This terahertz power is conveyed to the silicon structures reported in this work by means of 
a 2.8~mm-long linear-tapered spike that provides broadband index-matching, and is 
inserted directly into the hollow metallic waveguide. 
Alignment between these two components is performed with micrometer-controlled stages. 
Following transit through the silicon device, the terahertz waves are collected by a
second hollow metallic waveguide, and are demodulated 
via a mixer that is connected to a $\times$36 multiplier, which up-converts the 
microwave-range local oscillator signal of a spectrum analyzer. 
As a result, received terahertz power is viewed as a function of frequency in the
readout of the spectrum analyzer. 

Normalization by a reference measurement is required in order to convert received power 
into transmission magnitude. 
In the case of the coupling efficiency that is given in Fig.~\ref{fig:twolens}(b), 
a back-to-back measurement with the hollow waveguides coupled directly together 
(not shown) serves as reference. 
Thereafter, the negative decibel-value transmission efficiency is halved in order 
to estimate the coupling efficiency of a single lens. 
For the transmission of the DBR-based beam-splitter that is presented in 
Fig.~\ref{fig:DBR}(b), the two-lens sample shown in Fig.~\ref{fig:twolens}(a) is an 
appropriate reference measurement, as it facilitates isolation of the response 
of the optical filter itself. 
Finally, the liquid-sensing results given in Fig.~\ref{fig:liquid}(c) are normalized 
against the transmission response of the same device with the liquid removed. 
Additional reference measurements are taken between individual liquid-sensing experiments, 
in order to verify that there is no undesired residue on the sensing area that may 
obfuscate the results of subsequent tests. 

\subsubsection{Terahertz-range communications}

In order to support the demonstration of terahertz communications that is shown 
in Fig.~\ref{fig:comms}, a UTC-PD photomixer is excited with an on-off-keying-modulated
two-color infrared laser signal for which the beat frequency corresponds to the desired
output terahertz frequency. 
The resultant down-conversion produces terahertz waves that inherit the amplitude modulation 
of the optical signal. 
This terahertz power is accessed via a hollow metallic waveguide, which is coupled to 
Port~1 of the DBR-based frequency-division diplexer. 
Following transit through the silicon device, the modulated terahertz waves are collected
at either Port~2 or 3, and are input to an SBD-based detector, where they are 
converted directly to baseband via envelope detection. 
The demodulated bitstream is then amplified, clipped, and conveyed to both a 
bit error-rate tester and an oscilloscope simultaneously. 
The former serves to evaluate bit error-rate in real-time, and the latter displays 
the downconverted eye diagram. 

\note{The BER value corresponding to the FEC limit is selected based upon ITU-T 
recommendation G.975.1,\cite{union2004itu} which suggests that, for 7\% coding overhead, 
a pre-correction BER of $2\times10^{-3}$ serves as an upper-bound.
Above this BER value, error-correction is of limited utility to improve overall channel
quality.}

\subsubsection{Liquid handling}


All liquid chemicals are used as-delivered, and with the purity that is supplied 
by the manufacturer. 
Each liquid analyte is poured into an individual small glass petri-dish, to avoid 
cross-contamination. 
The petri-dish is placed upon a vertical translation stage, such that it may be brought 
into contact with the ATR-based liquid sensor without moving the device itself. 
This is necessary to maintain alignment with the hollow waveguides with which 
it is probed, with the aim being  that measurements in the presence of liquids are 
comparable with those without, for normalization purposes. 
Following probing, the stage bearing the sample is lowered, the petri-dish is removed, 
and the liquid is discarded.

\section*{Acknowledgments}

The authors wish to acknowledge support from the following grants: Core 
Research for Evolutional Science and Technology (CREST) program, Japan Science 
and Technology Agency (JST) (Grant No.~JPMJCR1534); Grant-in-Aid for Scientific 
Research, the Ministry of Education, Culture, Sports, Science and Technology of 
Japan (Grant No.~20H01064); the Deutsche Forschungsgemeinschaft 
(DFG, Geramn Research Foundation), Project ID ID287022738 TRR 196 (Project C06); 
the H2020-MSCA-ITN project TERAOPTICS- GA No.~956857.


\end{document}